\documentclass[draft]{agujournal}
\draftfalse

\journalname{Journal of Advances in Modeling Earth Systems (JAMES)}

\begin{document}

%
%

\title{Using machine learning to parameterize moist convection: potential for modeling of climate, climate change and extreme events}

%
%

\authors{Paul A. O'Gorman\affil{1},
John G. Dwyer\affil{1}\thanks{Current address: Dia\&Co, New York}}

\affiliation{1}{Department of Earth, Atmospheric and Planetary Sciences,
Massachusetts Institute of Technology, Cambridge, Massachusetts 02139, USA.}

\correspondingauthor{Paul A. O'Gorman}{pog@mit.edu}

\begin{keypoints}
\item Random-forest parameterization of convection gives accurate GCM simulations of climate and precipitation extremes in idealized tests 
\item Climate change captured when trained on control and warm climate, or only on warm climate, but not when trained only on control climate
\item Machine-learning parameterizations can also be interrogated to generate diagnostics of interaction of convection with the environment
\end{keypoints}

%
%


\begin{abstract} 
The parameterization of moist convection contributes to uncertainty in climate modeling and numerical weather prediction.  Machine learning (ML) can be used to learn new parameterizations directly from high-resolution model output, but it remains poorly understood how such parameterizations behave when fully coupled in a general circulation model (GCM) and whether they are useful for simulations of climate change or extreme events.  Here, we focus on these issues using idealized tests in which an ML-based parameterization is trained on output from a conventional parameterization and its performance is assessed in simulations with a GCM.  We use an ensemble of decision trees (random forest) as the ML algorithm, and this has the advantage that it automatically ensures conservation of energy and non-negativity of surface precipitation.  The GCM with the ML convective parameterization runs stably and accurately captures important climate statistics including precipitation extremes without the need for special training on extremes.  Climate change between a control climate and a warm climate is not captured if the ML parameterization is only trained on the control climate, but it is captured if the training includes samples from both climates. Remarkably, climate change is also captured when training only on the warm climate, and this is because the extratropics of the warm climate provides training samples for the tropics of the control climate.  In addition to being potentially useful for the simulation of climate, we show that ML parameterizations can be interrogated to provide diagnostics of the interaction between convection and the large-scale environment.
\end{abstract}

%
%

\section{Introduction}

General circulation models (GCMs) of the atmosphere and ocean are important
tools for climate simulation and numerical weather prediction. GCMs are based
on equations describing resolved dynamics (using the laws of conservation of
energy, momentum and mass) and parameterization schemes that represent subgrid
processes.  Parameterization schemes are necessary because there are
insufficient computational resources to resolve all relevant length and time
scales, but they are also the source of considerable uncertainties and biases
\citep[e.g.,][]{wilcox07,bechtold08,farneti11,stevens13}.  

One potential way forward is to use machine learning (ML) to create new
parameterization schemes by fitting a statistical model to the output of
relatively expensive physical models that more faithfully represent the subgrid
dynamics.  By minimizing the error between an ML model's predictions and the
known output over many training examples, ML models can learn complex mappings
without being explicitly programmed.  ML-based parameterizations have been
developed for radiative transfer
\citep[e.g.,][]{chevallier98,belochitski11} and for convective and
boundary-layer processes
\citep{krasnopolsky10,krasnopolsky13,brenowitz18,gentine18,rasp18}.  The use of
ML is also currently being explored for subgrid turbulence modeling for
engineering applications \citep[e.g.,][]{ling16,wang17}.  

In contrast to conventional parameterizations, an ML-based parameterization
takes a statistical approach and 
need not assume a simplified physical model such as the entraining plume that
is often used in convective parameterizations.  The resulting GCM is then a
hybrid model consisting of a physically-based component and one or more
ML-based components \citep{krasnopolsky_book}.  Such a hybrid approach is
particularly attractive if the most uncertain parameterizations in GCMs (which
often include many tunable parameters) can be replaced with ML-based
parameterizations that are training systematically.  An alternative approach to
leveraging high-resolution modeling or observations would be to use them to
optimize parameters while still retaining a physically-based subgrid model
\citep[cf.][]{emanuel99, schneider17}.

Subgrid moist convection is a good candidate for ML parameterization because
cloud-system resolving model (CRM) simulations are available to generate
training data, and because conventional parameterizations for moist convection
are responsible for considerable uncertainty in global modeling of the
atmosphere. Ideally a convective parameterization will accurately
represent the subgrid
fluxes of moisture, temperature and momentum associated with convective
instability, and account for both updrafts and downdrafts, mixing with the
environment, and cloud microphysical processes.  Historically a wide range of
approaches have been used to parameterize moist convection
\citep[e.g.,][]{arakawa04}.  Recent developments include efforts to include the
effects of the spatial organization of convection \citep{mapes11} and use of
super-parameterization in which CRMs are embedded in GCM grid boxes
\citep{khairoutdinov01,randall03}.  
Convective parameterizations affect the vertical structure of temperature and
humidity in the tropics \citep{held07, benedict13} and the ability of GCMs to
simulate the Madden Julian Oscillation and other tropical disturbances
\citep{kim12,benedict13}.  Convective parameterizations also strongly affect
how precipitation extremes are simulated \citep{wilcox07} and this helps to
explain the large spread in projected changes in precipitation extremes in the
tropics \citep{ogorman12}. 

CRM simulations differ from convective parameterizations in their predictions
for the response of convective tendencies to  perturbations in temperature and
moisture, both in terms of magnitude and vertical structure \citep{herman13}.
Furthermore, super-parameterization using embedded CRMs can reduce biases in GCM
simulations \citep[e.g.,][]{kooperman16}.  Thus, it is plausible that ML
parameterizations learned from CRM simulations could outperform conventional
parameterizations, and a GCM with an ML parameterization would be much faster
than a global CRM.

In a pioneering study, \citet{krasnopolsky13} used an ensemble of shallow
artificial neural networks (ANNs) to learn temperature and moisture tendencies
from CRM simulations forced by observations from a region of the equatorial
Pacific.  Tendencies from the resulting convective parameterization were
compared to tendencies from a conventional parameterization over the tropical
Pacific in a diagnostic test, but the key issue of fully coupling the ML-based
convection parameterization to the GCM was not addressed. Two recent studies,
published while this paper was in review, have found that a parameterization of
subgrid processes based on a shallow ANN ran stably in prognostic
single-column integrations when the
loss function included many time steps \citep{brenowitz18}, and that a
deep ANN trained on tendencies from a superparameterized GCM lead
to stable and accurate integrations in the same GCM \citep{rasp18}. 

Here, we use idealized tests to explore the potential of ML-based
parameterization for simulations of climate and climate change, and we demonstrate ways in which the ML-based parameterization can be used to gain physical
insight into the interaction of convection with its environment.  We train an
ML-based parameterization on the output of a conventional moist-convective
parameterization, the relaxed Arakawa-Schubert (RAS) scheme \citep{moorthi92}.
We then implement the ML-based parameterization in simulations with an
idealized GCM and compare the results to simulations with RAS.  This ``perfect-parameterization'' approach provides us with a simple testbed in which we can
cleanly investigate a number of important questions concerning how an ML-based
parameterization behaves when implemented in a GCM.  As described in detail
in \citet{moorthi92}, RAS is based on a spectrum
of entraining plumes and shares many features with real convection such as
sensitivity to humidity and temperature and nonlinear behavior such as only
being active under certain conditions.  
Since RAS is not stochastic and is
local in time and space, the idealized tests considered here may be viewed as a best-case deterministic
scenario for column-based machine learning that does not include the effect of
neighboring grid cells or past conditions.

We use a random forest (RF) \citep{breiman01,hastie_book} to learn the outputs
of the RAS convection scheme which are the convective tendencies of temperature
and specific humidity. Because we train on the output of RAS, the surface
precipitation rate is implied by the mass-weighted vertical integral of the
specific humidity tendency and does not have to be predicted separately. The RF
consists of an ensemble of decision trees, and each tree makes predictions that
are means over subsets of the training data. The final prediction from the RF
is the average over all trees. As described in the next section, the RF has
attractive properties for the parameterization problem in terms of preserving
physical constraints such as energy conservation, and we will show that it
leads to accurate and stable simulations of climate in the GCM. Running in a
GCM is a non-trivial test of an ML parameterization because errors in the
parameterization could push the temperature and humidity outside the domain of
the training data as the GCM is integrated forward
in time leading to large extrapolation errors
\citep[cf.][]{krasnopolsky08,brenowitz18}.  We also initially experimented with
using shallow ANNs (e.g., a single hidden layer with 60 neurons), but we found
that the resulting parameterization was less robust 
than the RF and did
not conserve energy without a post-prediction correction. 
We do not discuss these ANN results further given recent advances 
using different ANN training approaches and architectures
\citep{brenowitz18,rasp18}, and we instead focus on our promising results for
the RF parameterization.  

In addition to investigating the ability of the GCM with the RF
parameterization to accurately simulate basic statistics of a control climate,
we also investigate whether it accurately simulates extreme precipitation
events and climate change. For extreme events, we show that special training 
is not needed to correctly capture the statistics of these
events.  For climate change, we expect that an ML parameterization trained on a
control climate would not be able to generalize to a different climate to the
extent that this requires extrapolation beyond the training data.  Thus, the
extent to which generalization is successful can depend on both the magnitude
of the climate change and the range of unforced variability in the control
climate.  Interestingly, we also show that whether the climate is warming or
cooling is important, and that generalization across climates is related to
generalization across latitudes.  It is also important to know whether training
one parameterization on a combination of different climate states will work
well since this would be necessary for transient climate-change simulations.
Note that training on different climates is possible when training is based on
model output (e.g., from a global CRM) as long as these simulations can be run
for a sufficiently long period in a different climate.

Another promising aspect of ML is that it can be used to gain
insight from large datasets into underlying physical processes \citep[e.g.,][]{monteleoni13}. Here, we
explore whether the RF parameterization can be analyzed to provide insights
into the interaction of convection with the environment. We consider both the
linear sensitivity as has been previously discussed for moist convection
\citep{kuang10,herman13,mapes17}, and feature importance which is a common
concept in ML \citep{hastie_book} that does not require an assumption of small
perturbations.

We begin by describing the RF algorithm (section \ref{algorithm}), the RAS convection scheme and idealized GCM simulations used to generate training datasets (section \ref{ras_gcm}), and the training and validation of the RF convection scheme (section \ref{rf}). We discuss the ability of the idealized GCM with the RF scheme to reproduce the control climate including the mean state and extremes (section \ref{control_climate}), and its ability to capture climate change given different approaches to training (section \ref{climate_change}). We also show how the RF scheme can be used to provide insight into the importance of the environmental temperature and humidity at different vertical levels for convection (section \ref{feature_importance}). Lastly, we briefly discuss the ability of the RF scheme to represent the combination of the convection and large-scale condensation schemes (section \ref{lscale_cond}) before giving our conclusions (section \ref{conclusions}).

\section{Machine learning algorithm: random forest}
\label{algorithm}

A random forest (RF) is a machine-learning estimator that consists of
an ensemble of decision trees \citep{breiman01, hastie_book}.  
RFs are widely used because they do not require much preprocessing and 
they generally perform well over a wide range of hyperparameters.  The inputs to
the RF are referred to as features, and each decision tree is a recursive
binary partition of the feature space.  Each leaf of the tree contains a
prediction for the output variables that for continuous output variables is
taken to be the mean over the output from the training samples in that leaf.
Predictions of an RF are the mean of the predictions across all the trees,
and the purpose of having multiple trees  
is to reduce the variance of the
prediction since individual decision trees are prone to overfitting. 
The different trees are created by bootstrapping of the training
data and by only considering a randomly chosen subset of one third of the
features at each split when constructing the trees. The alternative approach
of considering all of the features at each split, referred to as bagging, gives
similar test scores for the problem investigated here. 

Training of the RF is an example of supervised learning in which an ML algorithm and a training dataset are used to learn a mapping between features and outputs
\citep[e.g.,][]{hastie_book}. The aim of the training is to minimize the mean squared error between the known and predicted outputs, and the resulting  model is referred to as a regression model because it predicts continuous variables.  Details of the features used and training of the RF are given in section \ref{rf}.

One major advantage of using an RF is that predictions are means over subsets
of the training data, and this leads to exact conservation of energy and
non-negativity of surface precipitation by the RF parameterization.  
Non-negativity of surface precipitation follows immediately since the
training samples all have non-negative precipitation, and the mean of a set of
non-negative numbers is a non-negative number.  To conserve energy in a
hydrostatic GCM, a convective parameterization that neglects convective
momentum transports should conserve column-integrated moist enthalpy, and this
is the case for RAS.  Moist enthalpy is a linear function of temperature and
specific humidity in our GCM, and thus the  predicted tendency by the RF of the
vertically-integrated moist enthalpy will be zero, ensuring energy
conservation.   One disadvantage of the RF is that 
considerable memory must be available when running the GCM in order to store the
tree structures and predicted values.  

The property that the RF predictions are averages over subsets
of the training data may also improve the robustness and stability of the RF
when implemented in the GCM. In particular, the predicted convective tendencies
cannot differ greatly from those in the training data, even if the RF is
applied to input temperature and humidity profiles that require extrapolation
outside of the training data (as can occur when an ML parameterization is
implemented in a GCM).

\section{Convection scheme and idealized GCM simulations}
\label{ras_gcm}

Our approach is to use a relatively complex convection scheme, typical of those used in current climate models, and implement it in an idealized GCM configuration to simplify the analysis of climate and climate change. The idealized GCM allows us to investigate the interaction between resolved dynamics and convection, but it does not include important complicating factors such as the diurnal cycle over land and cloud-radiation interactions.

For the convection scheme, we use the version of RAS that was implemented in the GFDL AM2 model \citep{anderson04}. This scheme is an efficient variant of the Arakawa-Schubert scheme \citep{arakawa74} in which the cloud ensemble is relaxed towards quasi-equilibrium. The basis of the scheme is an ensemble of entraining plumes that represent both shallow and deep convection. As discussed in \citet{held07}, the AM2 version of the scheme includes an entrainment limiter that is only active for deep convection. The inputs to RAS are the vertical profiles of temperature and specific humidity as a function of pressure, and the outputs are the tendencies of temperature and specific humidity. We do not consider convective momentum tendencies.

The idealized GCM is an atmospheric model based on a version of the GFDL spectral dynamical core coupled to a shallow thermodynamic mixed-layer ocean of depth 0.5m. There is no land or ice, and no seasonal or diurnal cycles. 
The GCM is similar to that of \citet{frierson06a} with the details as in \citet{ogorman08a} except that here we use the RAS convection scheme and we allow evaporation of falling condensate in the large-scale condensation scheme. The top-of-atmosphere insolation is imposed as a perpetual equinox distribution.  Longwave radiation is represented by a two-stream gray scheme with prescribed optical thickness as a function of latitude and pressure, and there are no water-vapor or cloud radiative feedbacks. The spectral resolution is T42, there are 30 vertical sigma levels, and the time step is 10 minutes.
The RAS scheme is responsible for most of the mean precipitation in the tropics, with the large-scale condensation scheme contributing to a greater extent at middle and high latitudes. This idealized GCM configuration (but with a simpler convection scheme) has previously been found to be useful for investigations of moist atmospheric dynamics and the response of precipitation to climate change \citep[e.g.,][]{ogorman09b,ogorman11,dwyer17}. 

The simulations with the RAS convection scheme are spun up over 700 days from
an isothermal rest state to reach statistical equilibrium. The simulations are
then run for a subsequent 3300 days, and this period is used to build training
datasets for ML as described in the next section.  Simulations with the
RF-based convection scheme are spun up over 700 days from the statistical
equilibrium state of the corresponding RAS simulation and then run for a
subsequent 900 days.  Simulations without any convection scheme are also used
for comparison purposes, and these are spun up over 700 days from an isothermal
rest state. All figures that present climate statistics are based on 900 days
at statistical equilibrium. The lower boundary condition and the
top-of-atmosphere insolation are zonally and hemispherically symmetric, and
thus differences between the hemispheres in figures are indicative of sampling
errors, except for the climate-change results in which the fields have been
symmetrized between the hemispheres to reduce noise.

We consider two climates: a control climate with a global-mean surface air temperature of 288K (similar to the reference climate in \citet{ogorman08a}), and a warm climate with a global-mean surface air temperature of 295K that is obtained by increasing the longwave optical thickness by a factor of 1.4 to mimic a large increase in greenhouse-gas concentrations. 

\section{Training and validation of the random forest}
\label{rf}

\begin{figure}[ht]
\centering
\includegraphics{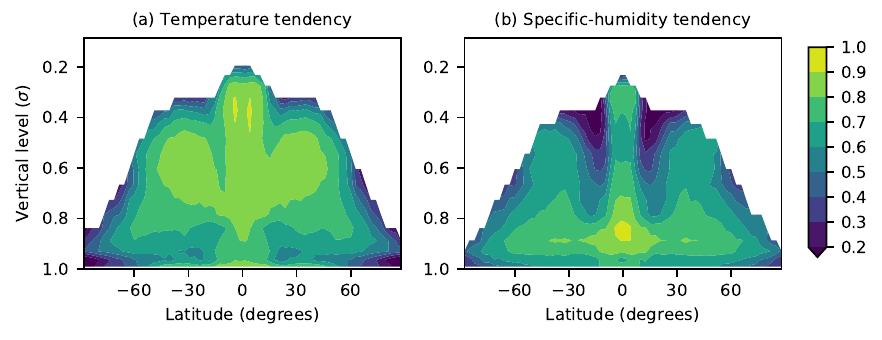}
\caption{Coefficient of determination R$^2$ for the convective tendencies from the random forest (RF) trained on RAS convective tendencies in the control climate for (a) temperature and (b) specific humidity. Results are plotted versus latitude and vertical level ($\sigma$) since the underlying GCM is statistically zonally symmetric.  R$^2$ is calculated based on the samples from the test dataset of the control climate (9900 samples for a given latitude and level), and it is only shown where the variance is at least 1\% of the mean variance over all latitudes and levels.}
\label{fig_latlev_Rsq}
\end{figure}

\begin{figure}[ht]
\centering
\includegraphics{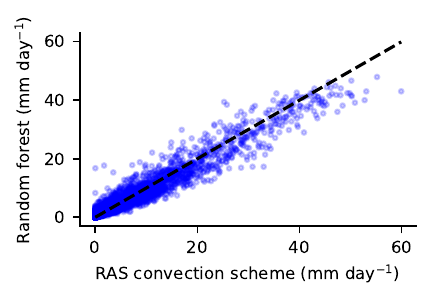}
\caption{Scatterplot of instantaneous precipitation from the RAS parameterization versus the random forest (RF) trained on the control climate. 
Precipitation is the negative of the  mass-weighted vertical integral of the specific humidity tendencies. 
The samples are from the test dataset for the control climate, and only a random subset of 10,000 samples are shown for clarity. The black dashed line is the one-to-one line. R$^2$ is 0.95 and the mean bias is negligible at $7 \times 10^{-5}$ mm day$^{-1}$.}  
\label{fig_precip_scatter}
\end{figure}

\subsection{Features and outputs}
The features are the inputs to the RF, and they
are chosen here to be the vertical profiles of temperature and specific
humidity (discretized at the vertical $\sigma$ levels) and the surface pressure.
Given that $\sigma$ is pressure normalized by surface pressure, these features are equivalent to the inputs to RAS which are the vertical profiles of temperature and specific humidity as a function of pressure.  Tests in which surface pressure is not included as a feature in the RF gave similar performance (note that the idealized GCM does not include topography).  We do not include surface fluxes as features since these are not an input to the RAS convection scheme.

The outputs are the vertical profiles of the convective tendencies of
temperature and specific humidity.  Cumulus momentum transports and
interactions of convection with radiation are not predicted since these are not
included in the idealized GCM.  The choice of output scaling for temperature
versus humidity tendencies affects how the RF fits the training data.  We chose
to multiply the temperature tendencies by the specific heat capacity of air at
constant pressure ($c_p$), and the specific humidity tendencies by the latent
heat of condensation ($L$) to give the same units for both tendencies. The
quality of the fit is similar if each output is instead standardized by
removing the mean and rescaling to unit variance. The training aims to
minimize the mean squared error summed over all the scaled outputs.

The nonlinear mapping that the RF learns may then be
written as $\mathbf{y} = f(\mathbf{x})$, where the vector of features is
${\mathbf{x} = (\mathbf{T}, \mathbf{q}, p_s)}$ and the vector of scaled outputs is
${\mathbf{y} = (c_p \partial \mathbf{T}/\partial t|_\mathrm{conv}, L \partial
\mathbf{q} /\partial t|_\mathrm{conv})}$. 
Here the vectors of temperature
and specific humidity at different vertical levels are denoted $\mathbf{T}$ and $\mathbf{q}$, respectively, and $p_s$ is the surface pressure. The time 
tendencies from convection are the output of the RAS convection scheme and are
denoted $\partial \mathbf{T}/\partial t|_\mathrm{conv}$
and $\partial \mathbf{q}/\partial t|_\mathrm{conv}$ for temperature and specific humidity, respectively. Since
convection is primarily active in the troposphere, we include the 21 $\sigma$-levels that satisfy $\sigma \geq 0.08$. Thus, there are 43 features and 42
outputs.  

We choose to have only one RF that predicts the convective tendencies of both temperature and specific humidity at all the vertical levels considered, and thus there are 42 outputs at each leaf of each tree. This column-based approach improves efficiency, and it ensures conservation of energy and non-negativity of precipitation as shown below. These two physical contraints would not hold if different RFs were used for predictions at each vertical level. Note also that we use the same RF for all latitudes (since RAS does not change depending on latitude) and that the RF is trained on data that includes both convecting and non-convecting gridpoints.

\subsection{Training and test datasets}

The temperature and specific humidity
profiles and surface pressure were output and stored from the GCM once a day
immediately prior to the point in the code at which RAS is called, and the
convective tendencies of temperature and specific humidity calculated by RAS
were also output and stored.  
We then randomly subsampled to 10 longitudes for a given time and latitude
to make the samples effectively independent. Noting
that the GCM is statistically zonally symmetric, time and longitude were
combined into one sampling index, and the samples were then randomly shuffled
in this index.  The first 70\% of the samples were stored for training while
the remainder of the samples were stored as a test dataset for model
assessment.  Lastly, the training and test datasets were randomly subsampled so
that the number of samples used at a given latitude is proportional to cosine
of latitude to account for the greater surface area at lower latitudes. (Not including the cosine latitude factor in sampling does not strongly affect the quality of the fit.) The training samples were then aggregated across latitudes and reshuffled, such that the final training dataset depends only on sample index and level.

\subsection{Fitting of Random Forest and choice of hyperparameters}

To train the RF, we use the RandomForestRegressor class from the scikit-learn
package version 0.18.1 \citep{scikit-learn}.  An advantage of the RF approach
is that there are only a few important hyper-parameters 
and they are relatively easy to tune. We analyzed the error of
the RF using 10-fold cross-validation on the training dataset from the control
climate. We varied the number of trees ({\it n\_estimators}), the minimum
number of samples required to be at each leaf node ({\it min\_sample\_leaf}),
and the number of training samples used ({\it n\_train}). 
Figures~S1, S2, S3 in the
Supporting Information show examples of the variations in error with these
hyper-parameters. Over the ranges shown, the error decreases with increasing
{\it n\_estimators} but the decrease in error is not very great for values
above $\sim$ 5 (Fig.~S1), the error decreases with increasing {\it n\_train}
but the decrease in error is not very great for values above $\sim 500,000$
(Fig.~S2), and the error is not very sensitive to {\it min\_sample\_leaf}
(Fig.~S3).  

The final choice of hyperparameters involves a tradeoff between the desire to
reduce error and the need for a fast parameterization that is not too large in
memory when used in the GCM.  In addition, we wanted to make sure that the size
of the training dataset would be feasible for generation by a high-resolution
convection-resolving or superparameterized model, with the caveat that training
on the output of such models may differ from what is described here.  Based on
these considerations and the error analysis discussed above, we chose to use
{\it n\_estimators=$10$}, {\it min\_sample\_leaf}=$10$, and {\it n\_train}=$700,000$. With the sampling approach described above, the training sample size is equivalent to just under 5 years of model output, but this could be reduced by sampling more often than once a day.

Using the above hyper-parameter choices, we fit RF models to the training
samples from the control RAS simulation, the warm RAS simulation, and the
combined training samples from the control and warm RAS simulations. In the
combined case, we still used 700,000 samples, and these were chosen after random shuffling the combined dataset.  The RF trained on the control simulation has an average number of nodes per tree of 62250, and it is 110Mb when stored as integers and single-precision floats in netcdf format for output to the GCM.  

\subsection{Validation on test dataset}

The performance of the RF for the control climate as evaluated based on the test dataset is shown in Figs.~\ref{fig_latlev_Rsq} and \ref{fig_precip_scatter}. Note that the RF was not trained on any of the samples from the test dataset. We use the coefficient of determination R$^2$ which is defined as one minus the ratio of the mean squared error to the true variance. R$^2$ for the tendencies of temperature and specific humidity is above $0.8$ in regions where the tendencies are large, such as the tropical mid-troposphere for temperature and the tropical lower troposphere for specific humidity, with generally higher R$^2$ for temperature as compared to specific humidity (Fig.~\ref{fig_latlev_Rsq}). The overall R$^2$ for the RF is $0.82$ as calculated over all test samples and levels, as compared to $0.86$ for the training dataset.
Note that the RF is specifically designed not to overfit the training data, in contrast to a single decision tree which could be trained to achieve perfect accuracy on a training dataset without achieving good performance on a test dataset.

The surface precipitation is also well captured by the RF (Fig.~\ref{fig_precip_scatter}) with an R$^2$ of 0.95 and a negligible mean bias of $7 \times 10^{-5}$ mm day$^{-1}$. The precipitation from RAS is the mass-weighted integral of the negative of the specific humidity tendency, and so precipitation does not require an additional prediction by the RF. Interestingly, the RF predictions of precipitation are reasonably accurate even at high values, and the ability of the RF to capture extremes of precipitation is discussed further in section \ref{control_climate}.

The RF trained on the warm climate does similarly well in predicting 
the test dataset of the warm climate (Fig.~S4) with 
R$^2$ of 0.77 for the tendencies and 0.93 for precipitation.
Issues of generalization and application to climate change are discussed
in section \ref{climate_change}.

\subsection{Conservation of energy and non-negative precipitation}

As can be seen in Fig.~\ref{fig_precip_scatter},
non-negativity of precipitation is ensured by the RF, and this holds
because the predictions of the RF are means over select training samples that all have non-negative precipitation. 
Conservation of the column-integrated moist enthalpy, which is linear in temperature and specific humidity in the GCM, is also ensured for the same reason.
The root-mean-squared error (RMSE) in conservation of column-integrated moist enthalpy in the control climate is very small at 0.2 W m$^{-2}$
for both the training dataset and the RF predictions on the test dataset.  
This error in conservation with the RF is substantially smaller than 
errors of order 50-100 W m$^{-2}$ that were reported
recently for ANN parameterizations \citep{brenowitz18,rasp18},
with the caveat that these reported conservation errors are not only due to errors in the ANNs and could be removed with a post-prediction adjustment.

\section{Implementation in GCM and simulation of control climate}
\label{control_climate}

\begin{figure}[ht]
\centering
\includegraphics{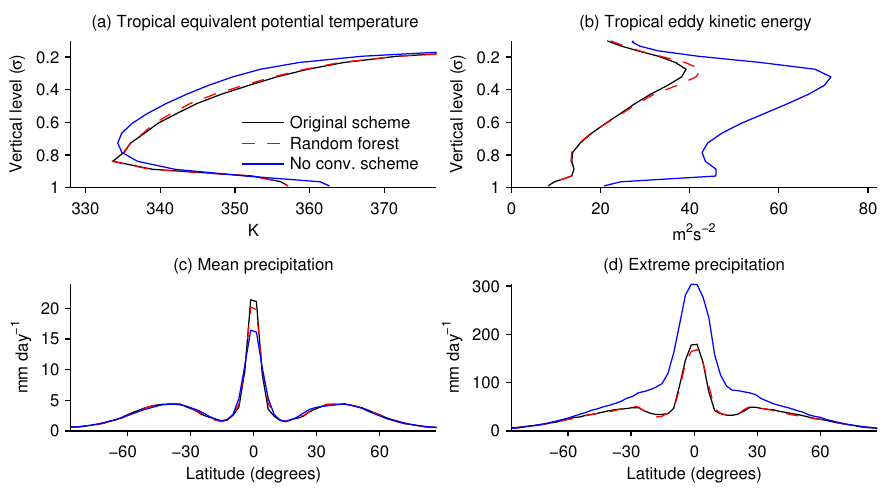}
\caption{Statistics from a GCM simulation of the control climate with the RAS parameterization (black) versus a simulation with the RF parameterization (red dashed) and a simulation without any convection scheme (blue). Shown are profiles of (a) tropical equivalent potential temperature versus vertical level ($\sigma$), (b) tropical eddy kinetic energy versus $\sigma$, (c) zonal- and time-mean precipitation versus latitude, and (d) the 99.9th percentile of daily precipitation versus latitude. Eddy kinetic energy is defined using eddy velocities with respect to the time and zonal mean.  The tropical equivalent potential temperature and tropical eddy kinetic energy are based on zonal and time means that are then averaged (with area-weighting) over 20$^\circ$S to 20$^\circ$N. 
}
\label{fig_control}
\end{figure}

We discussed the performance of the RF in offline tests in the previous
section. However, the most important test of a GCM parameterization is how it
performs in simulations with the GCM.  We consider the RF to be adequate as an
emulator for climate studies if GCM simulations with the RF can reproduce the mean climate and
higher-order statistics of simulations with the original parameterization. We
also compare to simulations without any convective parameterization to give a
benchmark for the magnitudes of any errors. 

Routines to read in the RF (stored as a netcdf file as discussed above) and to use it to calculate convective tendencies were added to the GCM which is written in Fortran 90. These routines simply replace the RAS convection scheme where it is called in the GCM. 
Introducing the RF-based parameterization into the GCM did not create any problems with numerical instability in the GCM simulations. The RF is faster than RAS by a factor of three.

The performance of the GCM with the RF in simulating the control climate is
shown in Fig.~\ref{fig_control}. Statistics are calculated using instantaneous
four-times-daily output for temperature, winds and humidity. Daily
accumulations are used for precipitation. The statistics shown are the
tropical-mean vertical profiles of equivalent potential temperature
($\theta_e$) (Fig.~\ref{fig_control}a) and eddy activity as measured by the
eddy kinetic energy (Fig.~\ref{fig_control}b), and the latitudinal
distributions of mean precipitation (Fig.~\ref{fig_control}c) and extreme
precipitation as measured by the 99.9th percentile of daily precipitation
(Fig.~\ref{fig_control}d).  In all cases, the GCM with the RF parameterization
correctly captures the climate as compared to the GCM with the RAS
parameterization (compare the black and dashed red lines in
Fig.~\ref{fig_control}).  This is particularly noteworthy for the
tropical $\theta_e$ profile, tropical eddy kinetic energy, and extreme
precipitation since these three statistics are sensitive to how convection is
parameterized and behave quite differently in simulations in which the
convection scheme is turned off and all convection must occur at the grid scale
(compare the black and blue lines in Fig.~\ref{fig_control}a,b,d).  
Snapshots of daily precipitation in Fig.~S5 illustrate that the RAS and
RF parameterizations result in weaker precipitation extremes and  more linear precipitation features in the intertropical convergence zone (ITCZ) as compared
to the simulations without a convection scheme.
Zonal and time-mean temperature is also well captured by the GCM
with the RF parameterization with a RMSE of 0.3K over all latitudes and levels.
The GCM with the RF parameterization generally does well for mean relative
humidity, although the values are slightly too
low in the tropical upper troposphere (Fig. S6).

Overall, these results suggest that the GCM with the RF parameterization can adequately simulate important climate statistics, including means, variances of winds (in terms of the eddy kinetic energy), and extremes. Climate statistics are the focus of this paper, but future work could evaluate the performance of the RF parameterization for other aspects such as wave propagation and initial value problems.

\section{Climate change and training in different climates}
\label{climate_change}

\begin{figure}[ht]
\centering
\includegraphics{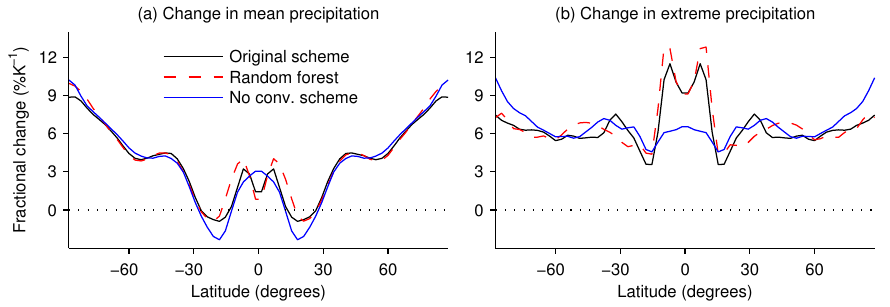}
\caption{Changes in (a) zonal- and time-mean precipitation and (b) the 99.9th percentile of daily precipitation between the control climate and the warm climate for simulations with the RAS parameterization (black), with the RF parameterization (red dashed) and with no convection scheme (blue). Changes are expressed as the percentage change in precipitation normalized by the change in zonal- and time-mean surface-air temperature. The changes in this figure have been averaged between hemispheres, and a 1-2-1 filter has been applied to reduce noise.}
\label{fig_precip_change}
\end{figure}

\begin{figure}[ht]
\centering
\includegraphics{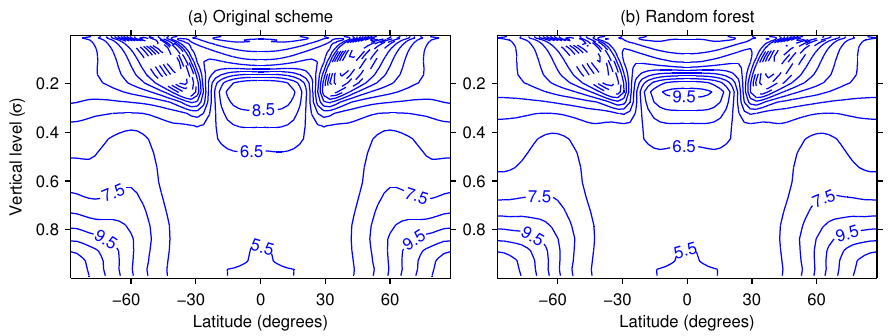}
\caption{Change in zonal- and time-mean temperature (K) versus latitude and vertical level ($\sigma$) between the control climate and the warm climate for simulations with (a) the RAS parameterization and (b) the RF parameterization. The contour interval is 1K and negative contours are dashed. The temperature changes have been averaged between hemispheres. The difference between results shown in (a) and (b)  over all latitudes and levels has a maximum absolute value of 1.1K and a root-mean-square value of 0.2K.
}
\label{fig_temp_change}
\end{figure}

\begin{figure}[ht]
\centering
\includegraphics{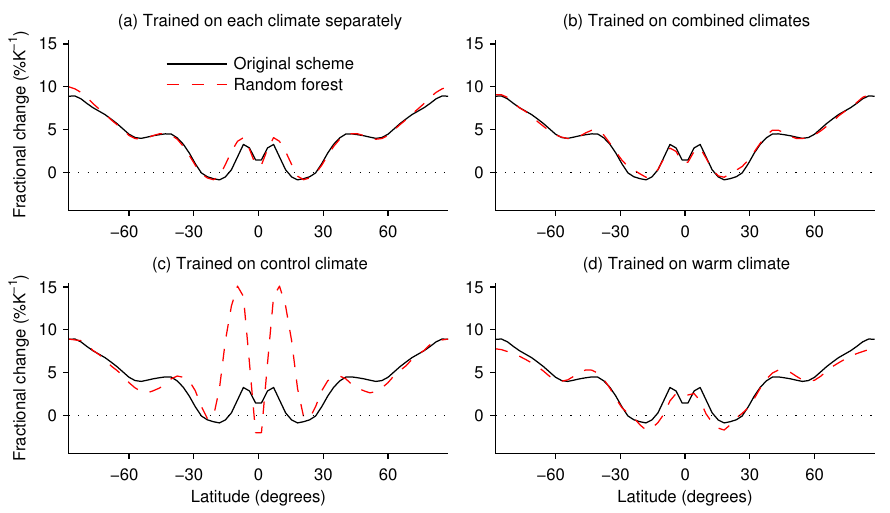}
\caption{Impact of training on different climates for the response to climate change of zonal- and time-mean precipitation for the RAS parameterization (black) and the RF parameterization (dashed red): (a) a different RF is trained for each climate separately, (b) one RF is trained using combined training data from both climates, (c), one RF is trained using training data from the control climate only, and (d) one RF is trained using training data from the warm climate only. Changes are expressed as the percentage change in precipitation between the control and warm climate normalized by the change in zonal- and time-mean surface-air temperature. The changes in this figure have been averaged between hemispheres, and a 1-2-1 filter has been applied to reduce noise.}
\label{fig_training}
\end{figure}

\begin{figure}[ht]
\centering
\includegraphics{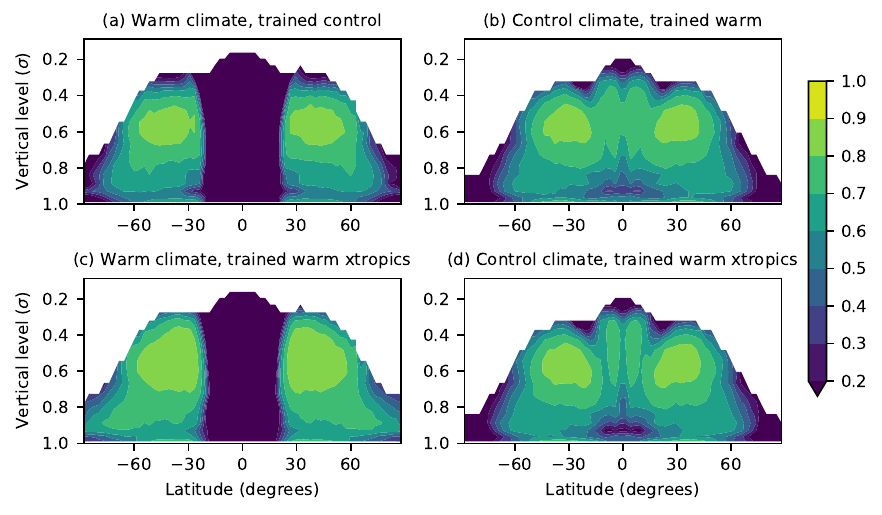}
\caption{Generalization of the RF to different climates or latitude bands as measured by R$^2$ of the convective temperature tendencies: (a) test dataset from warm climate with RF trained on control climate, (b) test dataset from control climate with RF trained on warm climate, (c) test dataset from warm climate with  RF trained on extratropics of warm climate, and (d) test dataset from control climate with RF trained on extratropics of warm climate.  
The extratropics is defined as latitudes poleward of 25$^\circ$ latitude in each hemisphere.  R$^2$ is only shown where the variance is at least 1\% of the mean variance over all latitudes and levels.
The ability of the warm-climate RF to predict the tropics of the control climate as shown in (b) comes from the ability of extratropical samples in the warm climate to predict the tropics of the control climate as shown in (d).}
\label{fig_generalization}
\end{figure}

We next assess the performance of the RF parameterization when applied to
climate change.  The RF introduces errors in simulating a given climate, and it
is important to quantify the impact of these errors on the simulated response 
to a forcing. In addition, it is interesting to know whether
an RF trained on a given climate can generalize to a different climate.

The most conservative approach is to train two RFs: one RF is trained on the control climate and used in a simulation of the control climate, and the other is trained on the warm climate and used in a simulation of the warm climate, and climate change is then calculated as the difference between the two climates.  
With this approach, the GCM with the RF parameterization accurately captures changes in climate, as shown for mean precipitation in Fig.\ref{fig_precip_change}a and extreme precipitation in Fig.~\ref{fig_precip_change}b. Note that RAS gives a strong increase in precipitation extremes in the tropics, albeit not as strong as when it was implemented in the GFDL coupled models CM2.0 and CM2.1 \citep{ogorman12}. The simulations in which the convection scheme is turned off have a more muted increase in extreme precipitation in the tropics. The GCM with the RF parameterization also faithfully captures the vertical and meridional structure of warming, with amplified warming in the tropical upper troposphere and polar amplification of warming in the lower troposphere (Fig.~\ref{fig_temp_change}). The GCM with the RF parameterization is slightly less accurate for the changes in mean relative humidity (Fig.S7), but these are generally small except near the tropopause where the upward shift in the circulation and thermal structure combines with sharp vertical gradients in relative humidity to
give large changes in relative humidity 
\citep[cf.][]{sherwood10, singh12}.

Next we assess the performance of different RF training approaches for climate
change as illustrated by the change in mean precipitation (Fig.~\ref{fig_training}).
Results for the vertical profile of warming in the tropics are shown in Fig.~S8
and lead to similar conclusions.  When an RF is trained separately for each
climate, the latitudinal profile of mean precipitation change is correctly
captured (Fig.~\ref{fig_training}a, repeated from
Fig.~\ref{fig_precip_change}a).  Training separately on different climates is
not necessarily feasible for simulations of transient climate evolution. An
alternative would be to train one RF using training data from a range of
different climate states. We test this approach here by combining training data
from the control and warm climates (but only using 700,000 training samples 
in total as before) and training one RF which is then used in GCM simulations of both climates. The GCM with the RF parameterization performs well in this case, as illustrated for the change in mean precipitation in Fig.~\ref{fig_training}b, and the control climate is also correctly simulated (not shown).

By contrast, training an RF on the control climate only and applying it in
simulations of both the control climate and the warm climate leads to
inaccurate climate-change results, with changes in precipitation in the tropics
and subtropics that are incorrect and much too large
(Fig.~\ref{fig_training}c). 
The RF fails to generalize because climate warming leads to higher temperatures and an upward shift of the circulation and
thermal structure \citep{singh12} including the tropopause \citep{vallis15},
but there are not examples in the training data from the control climate with
such high temperatures or such a high tropopause as occur in the tropics
of the warm climate. 
As a result, the vertical profile of tropical 
warming is severely distorted as shown in Fig.~S8c.
When the RF trained on the control climate is used to
predict the convective temperature tendencies from the test dataset for the warm
climate, it has no skill in the tropics equatorward of roughly
25$^\circ$ latitude (Fig.~\ref{fig_generalization}a). 
The cutoff latitude at which generalization fails may be estimated as the latitude at which the mean temperature in the warm climate is equal to the maximum mean temperature near the equator in the control climate. This estimate of the cutoff latitude is 19$^\circ$ for near-surface temperatures and 24$^\circ$ for temperatures at $\sigma=0.5$, which is comparable to what would be inferred
from Fig.~\ref{fig_generalization}a.  Note however, that errors in the convective tendences in the tropics are spread to other latitudes in the GCM simulations. 

The climate-change considered here is large (increase in global-mean surface temperature of 6.5K) and generalization
might be better for a smaller climate change. In addition, the control
simulation does not have a seasonal cycle or ENSO-like variability, both of
which might help by widening the range of training examples from the control
climate.  However, to the extent that an ML-based parameterization must
extrapolate at least at some times when applied to a warmer climate (e.g.,
during warm ENSO events), we expect it will not perform well.

Interestingly, training the RF on the warm climate and then applying it in
simulations of both the control and warm climates leads to good results for
climate change (Fig.~\ref{fig_training}d).  The vertical profile of tropical
warming is also well captured with a peak warming in the upper troposphere that
is only slightly too strong (Fig. S8d).  Convective tendencies from the test
dataset of the control climate are well predicted by the RF trained on the warm
climate except at polar latitudes where the tendencies are small in magnitude
(Fig.~\ref{fig_generalization}b).

Why is climate change better simulated when training on the warm climate rather
than the control climate? For a given latitude in the control climate
with a certain surface temperature and tropopause height, it is possible to
find training samples at higher latitudes in the warm climate with a similar
tropopause height and surface temperature.   Consistent with this argument, if
training of the RF on the warm climate is limited to samples from the
extratropics (latitudes poleward of 25$^\circ$ latitude in each hemisphere), it
fails to predict the tropics of the warm climate as expected
(Fig.~\ref{fig_generalization}c) but it still does a good job of predicting the
tropics of the control climate (Fig.~\ref{fig_generalization}d).  However, when
training is based on the control climate, it is not possible to find training
samples with a sufficiently high surface temperature and high tropopause that
are needed for the tropics in the warm climate
(Fig.~\ref{fig_generalization}a).

The asymmetry in the ability to generalize for climate cooling versus warming ultimately arises from differences between the tropics and extratropics. In the tropics, there is weak temperature variability and a warming climate quickly leads to problems for generalization. At high latitudes, moist convection is less important and there is more internal temperature variability which helps to broaden the range of training samples and makes it easier to generalize to a cooler climate. The meridional temperature gradient is also larger outside the tropics which means that different surface temperatures can be reached by moving a smaller distance in latitude.

Overall, our results show that the RF parameterization performs well in
simulations of climate change when the training data includes samples from both
climates. An RF can be trained separately for each climate or the training data
from both climates can be pooled to train one RF. Training on only the control
climate gives poor results as might be expected. However, 
training on only the warm
climate leads to remarkably good results for climate change,
and this is because a given latitude in the control climate can be predicted by a higher latitude in the warm climate.

\section{Feature importance of convection and sensitivity to perturbations}
\label{feature_importance}

\begin{figure}[ht]
\centering
\includegraphics{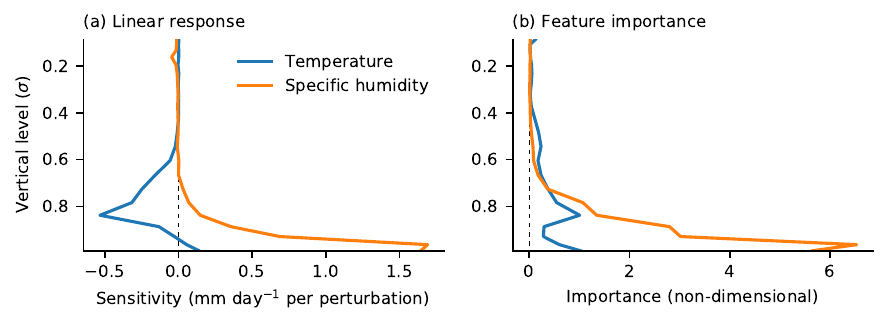}
\caption{Diagnostics measuring the responsiveness of convective tendencies to input temperature (blue) and specific humidity (orange) at different vertical ($\sigma$) levels according to the RF parameterization trained on the control climate. (a) Sensitivity of surface precipitation in mm day$^{-1}$ to perturbations in input temperature and specific humidity at different vertical levels ($\sigma$). The total perturbations are 1K for temperature and 1g kg$^{-1}$ for specific humidity and these are applied to samples with non-zero precipitation from the test dataset. The sensitivities have been rescaled to approximately represent the response to a 50hPa-deep input perturbation centered at a given level.  (b) Feature importance of input temperature and specific humidity at different $\sigma$ levels for convection (including both occurrence and strength of convection). The importance values have been rescaled by $1/d\sigma$ to account for the uneven $\sigma$ spacing. The vertically integrated importance is 0.24 for temperature and 0.75 for specific humidity (the importance of surface pressure is 0.01).}
\label{fig_linear_response_feature_importance}
\end{figure}

ML-based parameterizations could also be useful for building physical understanding about the interaction of convection with the large-scale temperature, humidity, and wind shear. Building an ML-based parameterization results in a nonlinear mapping that can be subsequently interrogated to learn about the underlying dynamics. We explore this possibility here in two different ways. First, we use the RF paramerization to generate a linear-response function for the response of convective precipitation to small perturbations in temperature, specific humidity and surface pressure.  Second, we use the concept of ``feature importance''  which seeks to measure the importance of the different input features (here temperature and humidity at different levels and surface pressure) for the RF predictions \citep[e.g.,][]{hastie_book}. The feature importance calculated here includes information on what features are important for both the occurrence and strength of convection, and it differs from the ``importance profiles'' discussed by \citet{mapes17} which we refer to here as linear response functions. 

For both the linear response function and the feature importance, we present
results for the RF trained on the control climate.  These results are
based on the RAS parameterization, and it would been possible to more directly
calculate the linear response function of RAS without the intermediate step of
the RF. However, if an RF is trained using high-resolution convection resolving
simulations, the RF mapping could be directly interrogated without the need to
run additional CRM simulations perturbed by forcings at different levels. And
as we will show, the feature importance is a useful additional diagnostic
for the interaction of convection with the large-scale environment.

\subsection{Linear-response function}

The linear-response function is similar to those that have previously
constructed for moist convection based on CRM simulations or convective
parameterizations \citep{kuang10,herman13,mapes17}. The input temperature,
specific humidity, and surface pressure of samples with non-zero precipitation
from the test dataset of the control climate (including all latitudes)
are systematically perturbed, and
the RF is applied to the unperturbed and perturbed samples.  For simplicity,
the resulting changes in the predicted tendencies from the RF are measured by
the perturbation in the predicted surface precipitation rate.  The
perturbations are added at each level and for each variable (temperature,
humidity, or surface pressure) separately.  The magnitude of the input
perturbation is $dT = 0.5$K for temperature, $dq=0.5$g kg$^{-1}$ for specific
humidity, and $dp_s=0.25$hPa for surface pressure. Both positive and negative
perturbations are used, and the reported sensitivity is the precipitation for
the positively perturbed input minus the precipitation for the negatively
perturbed input (thus representing the response to total perturbations of $1$K,
$1$g kg$^{-1}$ and $0.5$hPa) averaged across samples.  The $\sigma$ levels are
unevenly spaced, and so the results for temperature and specific humidity at a
given level are multiplied by $0.05/d\sigma$ to approximately represent the
response to a $50$hPa-deep input perturbation centered at that level.  Note
that the RF mapping is piecewise constant and not everywhere differentiable, but our perturbations are sufficiently large that this is not a
problem for estimating the linear response function, and we have confirmed
that the sensitivities are approximately doubled in size when the perturbations
are doubled in size.  

The linear response function is shown in
Fig.~\ref{fig_linear_response_feature_importance}a and is similar in some
regards to what was found by \citet{mapes17} based on CRM simulations of
unorganized convection in radiative-convective equilibrium at tropical temperatures.  Note however that our results are for a convective
parameterization in a full GCM simulation and including all latitudes; the
maximum absolute values are greater and more similar to what is found by
\citet{mapes17} if we only consider the equatorial region (not shown).  Surface
precipitation increases with moistening of the atmosphere, particularly at
lower levels. This sensitivity is consistent with the positive effect of
moisture on the buoyancy of a lifted parcel through both its initial
moisture and the effect of entrainment of environmental air. However, the
sensitivity to moistening is close to zero for levels above $\sigma=0.7$,
unlike what was found for CRM simulations by \citet{mapes17}, and possibly
indicative of a flaw that is common to convective parameterizations
\citep{derbyshire04}. Surface precipitation increases for a near-surface
warming, but decreases more strongly for a warming higher up, consistent with
the effect of warming on the buoyancy of a parcel lifted from near the surface.
For completeness, we note that the response to a surface pressure perturbation of 0.5hPa is 0.0015 mm day$^{-1}$.

\subsection{Feature importance}
Feature importance is shown in
Fig.~\ref{fig_linear_response_feature_importance}b based on the feature
importance metric that is implemented in RandomForestRegressor class of
scikit-learn (see \citet{hastie_book} for a more general discussion).  
For a given feature, this metric measures the total decrease in
mean-squared-error across nodes in a decision tree that split on that feature,
weighting by the fraction of samples that reach a given node, and then averaged
across trees in the ensemble. The
resulting importance values are normalized to sum to one over all features. To
account for the uneven spacing of the $\sigma$ levels, we multiply the feature
importance values for temperature and specific humidity by $1/d\sigma$.
Similarly to the results from the linear response function, the results from
the feature importance analysis imply that RAS precipitation is strongly
sensitive to low-level moisture and to temperature near $\sigma=0.8$. In
addition, we find that moisture is generally more important than temperature,
with the vertically integrated importance being 0.74 for specific humidity
versus 0.24 for temperature (and surface pressure is not important at 0.01). 

Advantages of feature importance compared to the linear response function
include that it doesn't require an assumption of small
perturbations and that it makes it easy to compare the importance of different
variables (e.g., humidity versus temperature).  Note that the linear response
functions for temperature and specific humidity are
not directly comparable because they assume a certain size of perturbation in
each variable and they have different units.  
It would also be possible to calculate feature importances for a classifier trained on the occurrence of convection to determine which features are most important for the occurrence of convection separately from the strength of convection.
On the other hand, the linear response function gives information on the
sign of the response, and the magnitudes of the sensitivities are easier to
interpret physically. Thus, both metrics are complementary and can
be used together to gain insight from the ML parameterization into
the interaction of convection with the environment.

\section{Replacing both the large-scale condensation and convection schemes}
\label{lscale_cond}

So far we have chosen to replace the moist convection scheme with an ML algorithm and to continue using conventional parameterizations for the large-scale condensation, radiation, and boundary layer schemes. When training on the output of high-resolution models it would be possible to either allow the ML algorithm to represent all of these schemes in a GCM or to use it for only some of them.
An advantage of replacing all of the schemes is that there can be significant compensation between tendencies from different schemes and there is not a clean physical separation of the different processes \citep[e.g.,][]{arakawa04}. However, it could be argued that some of the processes are easier to represent accurately with a conventional parameterization.

To explore this issue, we also tried replacing the sum of tendencies from the
moist convection scheme and the large-scale condensation scheme with an RF.
Using the same approach to training the RF as described above was found to give
poor results for relative humidity when the RF was implemented in the GCM
(compare Fig.~S6a and Fig.~S9a),
particularly in the extratropical upper troposphere where it can become
negative since the GCM does not enforce positive humidity. We
found through experimentation that the problem with the relative humidity was
largely eliminated by adjusting the training approach to take into account 
the properties of the large-scale condensation scheme (Fig.~S9b). We switched  to relative humidity as
the humidity input feature since large-scale condensation is sensitive to
saturation, we switched the output scaling of the tendency of specific humidity
to $L \sigma^{-3}$ instead of just $L$, to more strongly
weight the upper troposphere where large-scale condensation is
important for the relative humidity, and we removed the cosine latitude factor
in the sampling used to generate training and test datasets since large-scale
condensation is important at higher latitudes. 
All other aspects of the
training and parameters of the RF remain the same.

The RF with this choice of sampling, features and output scaling correctly
predicts the combined convective and large-scale condensation tendencies and
surface precipitation when applied to the test dataset (Fig.~S10 and S11), with
an overall R$^2$ for the tendencies of 0.83 and for the precipitation of 0.93.
When the RF is implemented in the GCM, replacing both the large-scale
condensation and moist convection schemes, it leads to  accurate simulations of
the control climate (Fig.~S12).  However, the precipitation response to climate
change is not accurate in the tropics (Fig.~S13), possibly as a result of the
need to simultaneously parameterize different processes at different vertical
levels, but it would be worthwhile to further explore the best choices of
features and output scalings for this case.

\section{Conclusions}
\label{conclusions}

We have investigated how an RF-based parameterization of moist convection
behaves when implemented in a GCM in an idealized setting. Encouragingly, the
RF parameterization was found to lead to robust and accurate simulations of the
control climate.  The use of a decision-tree-based approach made it
straightforward to ensure physical constraints such as energy conservation are
preserved by the parameterization.  Other approaches could be used to ensure
physical constraints are obeyed (such as adding an adjustment to the predicted
temperature tendencies to exactly conserve energy) but a decision tree approach
is attractive in ensuring they are exactly satisfied to the extent that they
hold in the training data.  The RF parameterization was also found to perform
well in the GCM in terms of simulation of extreme precipitation events, without
the need for specialized training on those events.

Climate change was accurately simulated when training samples from both the control and warm climate were used, and combining the training samples from both climates to train one RF was adequate. However, the RF trained in the control climate did not generalize to the warm climate, and the cutoff latitude at which it failed to generalize is approximately equal to the latitude at which the mean temperature in the warm climate is equal to the maximum mean temperature near the equator in the control climate.  Remarkably, training on just the warm climate gave good results for climate change. In effect, a given latitude in the control climate is predicted by samples from higher latitudes in the warm climate. The asymmetry between generalization for a warming versus a cooling climate relates to the weaker internal temperature variability, weaker meridional temperature gradients, and greater importance of moist convection in the tropics versus higher latitudes.

We have also illustrated how an ML parameterization can be interrogated to learn about underlying physical processes. First, the RF parameterization is useful as a means to efficiently generate linear response functions for small perturbations. Second, the RF parameterization can be use to measure the importance of different environmental variables such as temperature and humidity at different levels for convection, without the need to assume small perturbations. Feature importance could be further investigated separately for the occurrence of convection and the intensity of convection when it is occurring.

The setting we have used is idealized both in terms of using an aquaplanet GCM
and in terms of learning from a conventional parameterization rather than from
high-resolution simulations. Other studies have demonstrated that learning from
CRM simulations or superparameterized models is feasible
\citep{krasnopolsky13,brenowitz18,gentine18,rasp18}.  
When training on resolved
convection rather than a conventional parameterization, processing is needed
to calculate the appropriate convective tendencies to train on (e.g., through
coarse-graining approaches) 
and the interpretation of feature importance and
linear response functions are complicated by the presence of other
dynamical processes in addition to moist convection.  Some of the
interesting issues that remain to be explored include whether an ML
parameterization should be non-local in space and time, whether it should be
applied in addition to boundary-layer, radiation and large-scale cloud schemes
or replace all of these, and the extent to which convective-momentum tendencies
can be predicted.  Feature engineering, akin to our use of relative humidity
and a vertical weighting function in section \ref{lscale_cond}, is likely to be
useful in achieving good performance.  Extending to a more realistic GCM with
land brings up additional technical problems such as the strong diurnal cycle
over land and the need to predict convection at different elevations in the
presence of topography.  
These are non-trivial challenges, but our results
suggest that the use of ML is promising both for development of
new parameterizations and for new diagnostics of the interaction of subgrid
processes with the large-scale.

\acknowledgments
Seed funding for this research was provided by the MIT Environmental Solutions Initiative (ESI).  J.G.D. acknowledges support from an NSF AGS Postdoctoral Research Fellowship under award 1433290. P.A.O'G. acknowledges support from NSF AGS
1552195 and AGS 1749986. The scikit-learn package is available at http://scikit-learn.org/. The training and testing data, associated code, and RF estimators are available at zenodo.org \citep{ogorman18}. We thank two anonymous reviewers for their comments on the paper.


\listofchanges

\end{document}